\documentclass[nofootinbib,amsmath,amsfonts,amssymb,onecolumn,superscriptaddress,showpacs]{revtex4}
\usepackage{graphicx,psfrag,color}
\allowdisplaybreaks

\begin{document}

\hfill{UMD-PP- 014-015}

\title{Supernova Bounds on the Dark Photon Using its Electromagnetic Decay}
\author{Demos Kazanas}
\affiliation{NASA Goddard Space Flight Center, Greenbelt, MD, USA}
\author{Rabindra N. Mohapatra}
\affiliation{Maryland Center for Fundamental Physics and Department of Physics, University of Maryland, College Park, Maryland 20742, USA}
\author{Shmuel Nussinov}
\affiliation{School of Physics and Astronomy, Tel Aviv University, Tel Aviv, Israel}
\affiliation{Schmid College of Science,Chapman University, Orange, California 92866, USA}
\author{Vigdor L. Teplitz}
\affiliation{Physics Department, Southern Methodist University, Dallas}\affiliation{NASA Goddard Space Flight Center, Greenbelt, MD, USA}
\author{Yongchao Zhang}
\affiliation{Maryland Center for Fundamental Physics and Department of Physics, University of Maryland, College Park, Maryland 20742, USA}
\affiliation{Center for High Energy Physics, Peking University, Beijing, 100871, P. R. China}
\date{\today}

\begin{abstract}
  The hypothetical massive dark photon ($\gamma'$) which has kinetic mixing with the SM photon can decay electromagnetically to $e^+e^-$ pairs if its mass $m$ exceeds $2m_e$ and otherwise into three SM photons. These decays yield cosmological and supernovae associated signatures. We briefly discuss these signatures, particularly in connection with the supernova SN1987A and delineate the extra constraints that may then arise on the mass and mixing parameter of the dark photon. In particular, we find that for dark photon mass $m_{\gamma'}$  in the 5-20 MeV range, arguments based on supernova 1987A observations lead to a bound on $\epsilon$ which is about 300 times stronger than the presently existing bounds based on energy loss arguments.
  \end{abstract}
\maketitle

\section{Introduction and Motivation}
The detection of the $\sim{10}$ second neutrino pulse from Supernova 1987A, observations of its blue giant progenitor and of the subsequent electromagnetic signals have contributed a great deal to our understanding of supernova physics. Conversely new physics manifesting via neutrinos or new theoretically motivated particles is strongly constrained by these observations.  A path to many such constraints utilizes the fact that various stars do not suffer excessive energy losses via the production and emission of these new particles. As shown by many authors (for a review and references, see ref.~\cite{raffelt}), such considerations are particularly strong when we consider the very hot and dense core of the supernovae. Due to its high nuclear density, the core can, in the standard model (SM), cool only via emission of neutrinos. The collapse energy of a core of mass $M \sim 1.5 M_\odot$ and radius $R_c\approx 10 $ km is approximately $\frac{3}{5} \frac{ G_N M^2}{R_c} \sim (3-5) \times 10^{53}$ erg $\sim (2-4)\times 10^{59}$  MeV. The collapse heats up the core to a high temperature and generates a large flux of neutrinos of all six types (including neutrinos and antineutrinos) of average energy O($\sim 15$ MeV). Calculation of the mean free path for neutrino collisions suggests that the neutrinos will be trapped for a while in the core before diffusing out and the resulting neutrino emission based cooling is then expected to last for a few seconds.

The approximately twenty neutrinos observed from supernova 1987A~\cite{SN1987A-neutrino} confirm the expected pulse duration, it's energy spectrum and overall intensity. Hypothetical weakly coupled particles can -- for not too weak a coupling to electrons and nucleons -- be amply produced in the hot core and more readily escape, depleting the neutrino pulse and shortening its duration. This will contradict observations and thereby exclude significant domains of couplings of particles such as axions, KK recurrences of light particles, right handed neutrinos, super-light gravitinos etc, which can be thermally produced in the core and escape, for a wide range of masses and couplings. In this paper, we focus on the dark photon $\gamma'$ associated with a hidden $U(1)'$, which has been widely discussed in the literature in connection with attempts to understand the dark matter of the universe~\cite{dp-review}. These particles arise also in a class of particle physics models known as mirror models discussed in connection with neutrino physics as well as dark matter~\cite{mirror}. The dark photon is very weakly coupled to the SM particles and also expected to be light. Therefore, supernova observations should provide useful constraints on the properties of this particle.

In general dark photon models, gauge invariance allows a mixing of type $\epsilon F_{\mu\nu} F'^{\mu\nu}$, where $F_{\mu\nu}$ is the antisymmetric tensor made out of the SM $U(1)_Y$ gauge boson and $F'^{\mu\nu}$ is the corresponding one for the $U(1)'$ gauge field associated with the $\gamma'$. The primary channel through which $\gamma'$ couples to SM particles is through the kinetic mixing with photon with strength parameter $\epsilon$ given above. In the simple $U(1)'$ extensions of SM, $\epsilon$ is an arbitrary parameter, but in the context of a grand unified or other non-abelian embedding of either the SM or $U(1)'$, where, at the tree level, $\epsilon=0$, any super-heavy ``$Y$'' particles that carry both the SM $U(1)_Y$ and $U(1)'$ charges, can generate kinetic mixing $ \epsilon F_{\mu\nu} F'^{\mu\nu}$ at one or higher loop level. The mixing is then proportional to $ee'/{16\pi^2}$ and varies only logarithmically with the mass ratio $M_Y/M_{SM}$ with $M_{SM} \sim$ TeV to MeV the mass scales in the SM. Thus if $U(1)'$ is eventually unified with the SM gauge interactions, then we will have $e'\sim {e}$ and $\epsilon\sim 10^{-3}$ might be expected. However $U(1)'$ may have very small $e'$ charges leading to much smaller mixing, $\epsilon$. It is therefore of interest to consider a broad range of $m_{\gamma'}$ and $\epsilon$ in considering constraints on this new particle. Recent laboratory experiments have indeed already eliminated a wide range of $m_{\gamma'}$ for $\epsilon \geq 10^{-4}-10^{-5}$ or so. For smaller mixings, astrophysical observations such as from the supernova can be used and general energy loss arguments sketched above do indeed imply that $\epsilon < 10^{-10}$ for $m_{\gamma'} \leq 20$ MeV or so~\cite{dp-ex-SN1,dp-ex-SN2,dp-ex-SN3} (see Fig.~\ref{fig:constraint}). In this note we focus on the impact of subsequent decay of the $\gamma'$ on observations and discuss bounds from them on the $(m_{\gamma'},\,\epsilon)$  plane. For most of the paper, we will treat the $\gamma'$ on it's own as a particle with a mass $m_{\gamma'}$ and a mixing $\epsilon$ with the photon.

The $\gamma'$ produced in the supernova can decay to $e^+e^-$ or $3\gamma$'s depending on its mass. We will see that these decays will help to strengthen the bounds on $\epsilon$\footnote{It has been pointed out that stronger bounds on $\epsilon$ can arise from consideration of the effect of $\gamma'$ decays on Big Bang Nucleosynthesis~\cite{dp-cosmic}. These bounds arise in a different mass range of the dark photon and are complementary to our results.}. The boundaries of the region in the $(m_{\gamma'},\,\epsilon)$ plane excluded by the energy loss argument correspond to situations where the dark photons carry out of the core a sizable fraction of the huge collapse energy $W_{\rm total} \sim (3-5) \times 10^{53}$ ergs. Our main point is that if some fraction of this energy also manifests electro-magnetically, then the signature would be hard to miss. The fact that the optical, UV, X-ray etc signatures started only $\sim$ 3 hours after the collapse with the expected light-curve magnitudes and durations, extends further the disallowed $(m_{\gamma'},\,\epsilon)$ regions. Some of the arguments here are similar to those used in deriving limits on radiative decays of neutrinos from the supernova leaving the progenitor star~\cite{Dar,SN-nu1,SN-nu2} or the axion-photon conversion in the strong external fields \cite{Grifols}.

The paper is organized as follows: in sec. 2, we discuss preliminaries on the production and decays of the dark photon; in sec. 3, we consider the constraints when the $\gamma'$ decays outside the core but before it reaches the surface of the progenitor; in sec. 4, we discuss the same for the case when the $\gamma'$ decays to $e^+e^-$ in the vicinity of the progenitor star, namely at distances between $R_\ast$ (the radius of the star) and $2R_\ast$; in sec. 5, we discuss the case with dark photon mass below $2m_e$ and the resulting constraints on $\epsilon$ for this kinematic range. We conclude with a summary of our results in sec. 6.

\section{Preliminaries on $\gamma'$ production and decays}
In this section, we briefly discuss the production of the dark photon in the supernova core and its decay properties.

\subsection{Production}

In the supernova core, the dominate production process of the dark photons are the nucleon bremsstrahlung scattering $pp \rightarrow pp \gamma'$ and $pn \rightarrow pn \gamma'$, with the dark photon coupled to proton via the kinetic mixing $\epsilon F_{\mu\nu} F^{\prime\mu\nu}$. Following closely Ref.~\cite{dp-ex-SN2}, we calculate the energy emission rate of dark photon per unit volume from both the two bremsstrahlung processes,
\begin{eqnarray}
\label{eqn:QA'1}
Q_{\gamma'} = \int d\Pi_5 \ S \sum_{\rm spins}|\mathcal{M}|^2(2\pi)^4 \delta^4(p_1+p_2-p_3-p_4-p_{\gamma'}) E_{\gamma'} f_1f_2,
\end{eqnarray}
where $d\Pi_5$ is the phase space of the five incoming and outing particles, $S$ symmetry factor, $\mathcal{M}$ scattering amplitudes, $f_{1,\,2}$ the non-relativistic Maxwell-Boltzmann distributions of the two incoming nucleons in the non-degenerate limit. See the appendices in Ref.~\cite{dp-ex-SN2} for the calculation procedure and details. After some straightforward simplification, we find that, compared to $pn$, the $pp$ scattering process is highly suppressed by the ratio $m^2_{\gamma'}/\omega^2$ ($\omega$ being the momentum transferred in the $NN$ scattering), the symmetry factor $S$ and powers of the nucleon-pion coupling ratio $f_{pn}/f_{pp}$, thus in the calculation below we consider only the former one, with the energy emission rate expressed as\footnote{Some minor errors in the calculation of~\cite{dp-ex-SN2} have been corrected.}
\begin{eqnarray}
\label{eqn:QA'2}
Q_{\gamma'} &=& \frac{ f^4 e^2 \epsilon^2 \alpha_{\pi}^2 n_B^2 T^{2.5} }{8\pi^{1.5}m_N^{3.5}} \int dudvdzdx
\sqrt{uv} {\rm{e}}^{-u} \sqrt{1-\frac{q^2}{x^2} }\delta(u-v-x) \mathcal{I}_{pn} \,,
\end{eqnarray}
where $f \simeq 1$ is the pion-nucleon coupling, $e$ the electric charge, $\alpha_{\pi} \equiv (2m_N/m_\pi)^2/4\pi$ with $m_N$ and $m_\pi$ the masses of nucleon and pion, $n_B = 1.2 \times 10^{38}/{\rm cm}^3$ and $T = 30$ MeV the baryon density and temperature in the supernova core, $u,\, v,\, z,\, x,\, q$ dimensionless parameters relating the masses, momenta and temperature in the scattering processes. $\mathcal{I}_{pn}$ is the integrand for the $pn$ process as function of the dimensionless parameters. Both the dimensionless parameters and the integrand are defined in the appendix.

In the absence of the dark photon decay, the integral is solely determined by the dark photon mass $m_{\gamma'}$. To obtain the luminosity due to the dark photon emissions, we have to integrate over the whole volume of the supernova core $V_c = \frac43 \pi R_c^3$, assuming simply the production rate is a constant within the core $R_c \simeq 10$ km. For a decaying dark photon, we have to include the decaying factor $\exp[-R/(c\tau_{\gamma'})]$ in the integral, which depends both on the mixing and mass parameters. We would also like to mention that when calculating the number emission rate per unit volume, the explicit factor $E_{\gamma'}$ in Eq.~(\ref{eqn:QA'1}) should be removed,
\begin{eqnarray}
\label{eqn:NA'}
N_{\gamma'} = \int d\Pi_5 \ S \sum_{\rm spins}|\mathcal{M}|^2(2\pi)^4 \delta^4(p_1+p_2-p_3-p_4-p_{\gamma'}) f_1f_2 \,.
\end{eqnarray}
In obtaining the following constraints, the equations~(\ref{eqn:QA'1}) and (\ref{eqn:NA'}) are the basics for the ``exact'' numerical calculations, with the decaying factor being also a crucial factor.

For our discussion below, we will take the total number of dark photons to be such that it carries away energy less than the luminosity corresponding to neutrinos. As a rather rough estimation, this corresponds to the value $\epsilon \sim 10^{-10}$ with each $\gamma'$ having an average energy of $\sim 20$ MeV. From the fact that the total energy loss is $(2-3)\times 10^{53}$ ergs, we deduce this average number to be $\sim 0.5\times 10^{58}\epsilon^2_{10}$ where $\epsilon_{10}= \frac{\epsilon}{10^{-10}}$.

\subsection{Decay}
\begin{figure}[t]
  \centering
  \includegraphics[width=0.37\textwidth]{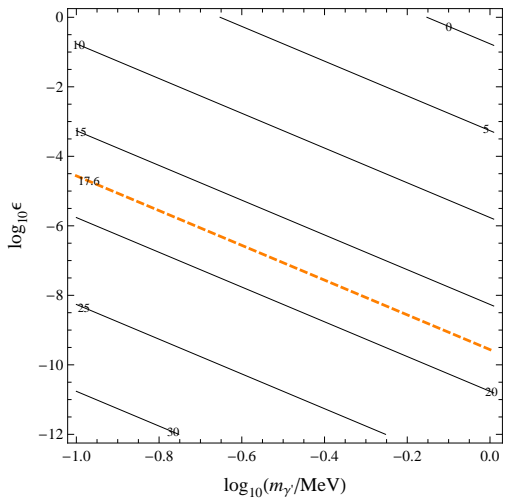}
  \vspace{-.3cm}
  \caption{The lifetime of dark photon $\log_{10} (\tau_{\gamma'}/{\rm sec})$ as functions of the mass and kinetic mixing parameters. The dashed line corresponds to the present age of our universe.}
  \label{fig:lifetime}
\end{figure}

If $m > 2m_e\sim$ { MeV}, the decay channel $\gamma'\rightarrow{e^+ e^-}$ is allowed and the decay rate is given by
\begin{eqnarray}
\Gamma (\gamma'\rightarrow e^+ e^-) = \frac13 \alpha \epsilon^2 m_{\gamma'}
\left( 1-\frac{4m_e^2}{m_{\gamma'}^2} \right)^{1/2}
\left( 1+ \frac{2m_e^2}{m_{\gamma'}^2} \right) \frac{m_{\gamma'}}{E_{\gamma'}},
\end{eqnarray}
where we have included the time dilation factor $E_{\gamma'}/m_{\gamma'}$. This leads to a decay length,
\begin{equation}
\label{eqn:decay}
L_{\rm decay} (\gamma'\rightarrow e^+ e^-) \sim 10^{12} \frac{E_{\gamma'}}{m^2_{\gamma'}} \left(\frac{1}{\epsilon_{10}}\right)^2 \, {\rm cm},
\end{equation}
where we have defined $\epsilon_{10}\equiv \frac{\epsilon}{10^{-10}}$ as before and $E_{\gamma'}$ and $m_{\gamma'}$ are expressed in MeV units. Once $m_{\gamma'} < 2m_e$, only the 3 photon decay mediated via an electron box diagram is allowed and the decay rate is given by~\cite{photon-splitting} (again including the time dilation factor):
\begin{eqnarray}
\label{eqn:3gamma}
\Gamma(\gamma'\to 3\gamma)\simeq \left(\frac{2\alpha^2\epsilon}{45}\right)^2\frac{1}{6\times 2^{11}\pi^9}\left(\frac{m_{\gamma'}}{m_e}\right)^8 m_{\gamma'} \cdot \frac{m_{\gamma'}}{E_{\gamma'}},
\end{eqnarray}
and the decay length then dramatically increases to
\begin{equation}
L_{decay} ( \gamma' \rightarrow 3 \gamma)
\simeq 10^{28}\epsilon^{-2}_{10} \left(\frac{\rm MeV}{m_{\gamma'}}\right)^{-10}
\left(\frac{E_{\gamma'}}{20\,{\rm MeV}}\right) \, {\rm cm}.
\end{equation}
For convenience we present in Fig.~\ref{fig:lifetime} the contours in the $(m_{\gamma'},\,\epsilon)$ plane of given decay length for the case of $3\gamma$ decay.

\begin{figure}
  \centering
  \includegraphics[width=0.5\textwidth]{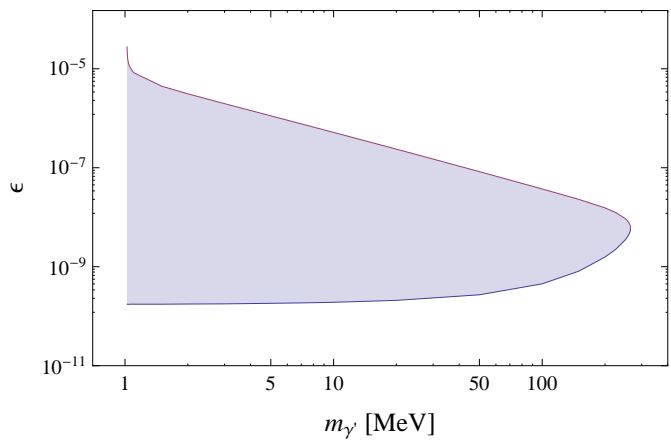}
  \vspace{-.3cm}
  \caption{The shaded region in the $m_{\gamma'} - \epsilon$ plane is excluded due to the simple luminosity argument.}
  \label{fig:constraint}
\end{figure}

Before presenting our new limits from the electromagnetic decays, let us start by reviewing the usual argument that the luminosity due to massive dark photon emission should not exceed that due to neutrino emission i.e. $L_\nu \simeq 10^{53} \, {\rm erg/s}$. As has already been noted in~\cite{dp-ex-SN2}, this excludes the shaded region shown in Fig.~\ref{fig:constraint}. To get the boundaries of the excluded region, we plug the decaying factor $\exp[-R_c \Gamma (\gamma'\rightarrow e^+e^-)]$ in Eq.~(\ref{eqn:QA'2}) and require that the total energy emission rate due to dark photons $Q_{\gamma'}V_c < L_\nu$. For a mixing parameter with large enough values, the observed luminosity $L_\nu$ sets a lower bound on the excluded region. On the other hand, when $\epsilon$ is very large, the dark photons produced would be trapped and decay in the core, not contributing effectively to the supernova cooling, which leads to the top edge in the figure.

\section{Limits when the $\gamma'$ decays inside the mantle}
If the $\gamma'$'s decay inside the mantle, the numerous and energetic resulting electrons and positrons can blow away all  the stellar material beyond the decay radius generating a fast moving hot ejecta starting an intense light emission earlier than the observed three hours delay. Demanding that this does not happen sets new strong limits on $\epsilon_{10}$.

\begin{figure}[t]
  \centering
  \includegraphics[width=0.5\textwidth]{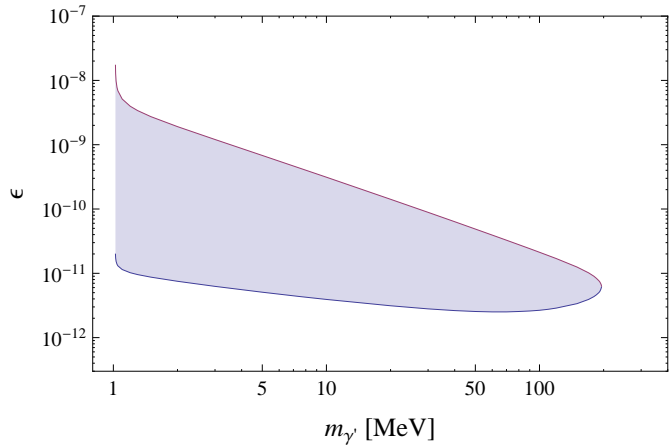}
  \vspace{-.3cm}
  \caption{The shaded region in the $m_{\gamma'} - \epsilon$ plane is excluded by the blow-away arguments.}
  \label{fig:constraint_mantle}
\end{figure}

The mantle at $r>R_0$ can be blown-off if the energy in the $e^+ e^-$ deposited therein by decays of the out-streaming $\gamma'$'s  exceeds the gravitational binding of the mantle. The mantle mass is  $\delta M=\int_{R_0}^{R_*} {r^2\rho(r)dr} $ (where $R_* = 3\times 10^{12}$ cm is the radius of the progenitor star), and the energy deposited is estimated to be $0.2\epsilon^4_{10}3.10^{53}$ ergs  (one $\epsilon^2$ factor reflects the reduced $\gamma'$ outflow from the core and another $0.2 \epsilon^2$ approximates the fraction $(R_*-R_0)/{L_{decay}}$ of the $\gamma'$ decays which occur inside the mantle. The blow-up condition then becomes
\begin{eqnarray}
0.2 \epsilon^4_{10} 10^{53}~{\rm erg} \gtrsim \frac{G_N M_\ast \delta M}{R_*} \,,
\end{eqnarray}
where $M_\ast$ is the star mass. The mass density of the progenitor $\rho(r)$ is monotonically decreasing as we move outward from the core but the density profile $\rho(r)$ of the progenitor is not well known and is model dependent, e.g. a polytrope model~\cite{polytrop} is often invoked. We therefore estimate $\delta M \sim 0.1 M_{\odot}$ for the shell including the last $20\%$ of the stelar radius (hence the 0.2 above). Assuming further that $M_\ast \sim 10 M_{\odot}$ and $R_\ast \sim 3\times 10^{12}$ cm, we find $\epsilon^4_{10} \lesssim 10^{-5}$. Since the decay involves both the mass $m_{\gamma'}$ and $\epsilon_{10}$, a more precise rendering of the bound is via a region in the $m_{\gamma'}-\epsilon_{}$ plane ruled out by the fact that the blow-up and its dramatic visual manifestation did not happen. In more exact numerical calculation, we plug in the energy emission rate $Q_{\gamma'}$ the factor
\begin{eqnarray}
\exp[-0.8R_\ast \Gamma(\gamma'\rightarrow e^+e^-)] - \exp[-R_\ast \Gamma(\gamma'\rightarrow e^+e^-)] \,,
\end{eqnarray}
which accounts for the dark photons decaying within the radius ($0.8R_\ast$, $R_\ast$). Assuming conservatively the emission duration $\Delta t = 1$ sec, and requiring that the energy transferred from the $e^+e^-$ pairs to the outer layer
\begin{eqnarray}
Q_{\gamma'} V_c \Delta t < \frac{G_N M_\ast \delta M}{0.8R_*} \,,
\end{eqnarray}
we get the constraint presented in Fig.~\ref{fig:constraint_mantle}. Note the improvement of the bound for higher masses.

We wish to note that we could have utilized the extended ($ 60 \times 10^3$ Km $= 3\times 10^{-3}R_\ast$) atmosphere of the Sanduleak progenitor with total column density of $\sim{10^6}$ gr/cm$^2$ in order to make similar estimates. Clearly in this case there would be no doubt that even a relatively small fraction of the collapse energy $10^{53}$ would have blown the atmosphere away. Our reasoning for using instead a nominal $0.1$ solar mass layer is that tenuous layers under the surface of the star with gravitational binding significantly smaller than the deposited energy cannot efficiently process all that energy into observable photons. This is similar to the case where a tiny fraction of dark matter particles in the supernova core cannot stop the outward flow of the $\gamma'$s. As we argue above, the density profiles of stars make our choice of $\delta M\sim 0.1 M_\ast$ quite plausible.

\section{Limit from decay near the outside of the progenitor's surface}
In this section, we focus on the constraints resulting from the decay of the dark photons near the surface, outside but not very far from the progenitor star. Similar constraints coming from the same observed lack of a prompt electromagnetic signal have been used in the past by A. Dar and S. Dado to constrain massive radiatively decaying neutrinos~\cite{Dar,SN-nu1}. We believe however that our work is the first utilizing these constraints in the dark photon context.

Using the decay length formula in Eq.~(\ref{eqn:decay}), with for $E_{\gamma'}/{m_{\gamma'}^2}\sim {1}$ (with $\gamma'$ energies and masses expressed always in MeV units) and $\epsilon_{10}=1$ we find that $L_D \sim 0.3 R_*$ where $R_* = 3\times 10^{12}$ cm is the radius of the Sanduleak Blue Giant progenitor star.

By far the most dramatic electromagnetic signal occurs when the parameters -- $m_{\gamma'} = 3$ MeV, average dark photon energies $E_{\gamma'}=25$ MeV and $\epsilon_{10}=1$ -- are such that the decay distance match the stellar radius $ L_{decay} = R_*= 3\times 10^{12}$ cm. In this case $e^{-R/R_\ast}-e^{-2R/R_\ast}= e^{-1}-e^{-2}\sim {20 \%}$ of the decays occur outside but within a distance $R_*$, namely inside the outer shell: $2R_* >r >  R^*$. Then $20\%$ of the energy carried by the $\gamma'$ which for $\epsilon_{10}=1$ is $ W({\rm tot})_{\epsilon=10^{-10}} \sim {W({\rm Collapse})} = 3\times 10^{53}$ ergs will be deposited in this region by the decays $\gamma'\rightarrow{e^+ e^-}$. As we shortly demonstrate all this energy will manifest finally via $\sim 1/2$ MeV photons so that $N_{\gamma}({\rm tot})= 10^{59}$. The total fluence at earth, $50 \, {\rm kpc} = 1.5 \times 10^{23}$ cm away from the SN, would then be $3.10^{11}{\rm cm}^{-2}$. The SMM satellite \cite{SN-nu1} established an upper limit on the fluctuation above background in this energy range of $0.1$ photons/ cm$^2$/sec. This is smaller by about 10 or 11 orders of magnitude compared to our prediction. This leads to a bound on $\epsilon$:
\begin{equation}
0.2\epsilon^4_{10}\leq 10^{-10}-10^{-11}\\ ~~{\rm or}~~ \epsilon_{10} \leq 5\times 10^{-3}-3\times 10^{-3}
\end{equation}
Note however that
for much smaller $\epsilon$ the decay lengths $\sim{R_*/\epsilon^2_{10}}$ are much larger than $2 R_*$.
It is still optimal to focus even in such cases on decays occurring inside the above $(R_*, 2R_*)$ region for the following reasons:

a) It takes 100 sec to traverse this region and the photons emerging from it, arrive promptly during a time interval of
$\sim{100~ {\rm sec}}/{\gamma^2_L}$ where $\gamma_L=E_{\gamma'}/m_{\gamma'}$ is the Lorentz factor of the $\gamma's$. Clearly this will make the signal far more intense and dramatic.

b) Even for $\epsilon^2_{10}\ll{1}$ a fraction of $\sim 0.2 \epsilon^2_{10}$ of all $\gamma's$ will still decay early within this near-by region. The density of the resulting $e^+ e^-$ produced by $\gamma'$ decays in there scales as $\epsilon^4_{10}$ (another factor of $\epsilon^2_{10}$ from the total number of dark photons produced in the core) only and not as $\epsilon^6_{10}$ as the overall density of the electrons and positrons produced by most of the decay within $R_*/{\epsilon^2_{10}}$. In particular the optical depth even for travel across the $R_*/{\epsilon^2_{10}}$ big sphere is reduced by $\epsilon^6_{10}$ versus just by $\epsilon^4_{10}$ as in the case considered here where only the effect due to the near-by decays is considered.

If for $\epsilon^2_{10}=1$ indeed $20\%$ of the $10^{58}\gamma's$ decay in the region considered the resulting $e^+e^-$ density is:
\begin{equation}
 n_{\gamma'}= \frac{ 2.10^{58} } { \frac{4\pi}{3} (6\times10^{12}{\rm cm})^3 } =  10^{19}~ {\rm cm}^{-3},
\end{equation}
with the annihilation cross-section of
\begin{equation}
\sigma_{ann}=  10^{-25}/{E_{\rm MeV}^2}\sim  10^{-28} {\rm cm}^2,
\end{equation}
where $E_{\rm MeV}$ is the center of mass energy of the annihilating $e^+e^-$. The optical depth accumulated over a generic distance of $L= R_*=3.10^{12}$ cm, then is:
\begin{equation}
\kappa= R_*n\sigma_{ann}= 3.10^3.
\end{equation}
Hence all the positrons will annihilate and we have a relativistic plasma which expands and cools as in the classic fire-ball model for a spherical intense $\gamma$ ray burst\cite{pac}.
The total fluence reaching earth at a distance $D=50$ kpc away during less than 100 seconds after the neutrino signal would then be  be $3.10^4 {\rm erg}/{\rm cm}^2$ and as discussed above the flux of $\sim {1/2}$ MeV photons would exceed the SMM limits by $\sim 10^{12}$.

As $\epsilon$ decreases the intensity of the $\gamma$ flash decreases for the following three independent reasons:

i) An $\epsilon_{10}^2$ factor suppresses the intensity of the $\gamma'$ burst from the SN core.

ii) The $\gamma'$ lifetime is prolonged by $\epsilon^{-2}_{10}$ and only a fraction $\epsilon^2_{10}$ of the $20\%$ of the gamma's that when $\epsilon^2_{10}=1$ decay between $R_*$ and $2 R_*$ will decay in there for the smaller $\epsilon^2_{10}$.

iii) The  optical depth then decreases by $\epsilon^4_{10}$ and the probability of annihilation is no longer 1 but Min$(1, 3.10^3\epsilon^2_{10})$. The last expression simplifies to just $3\times 10^3\epsilon^2_{10}$ for $\epsilon_{10}\leq{1/{60}}$ -- the region targeted.

\begin{figure}
  \centering
  \includegraphics[width=0.5\textwidth]{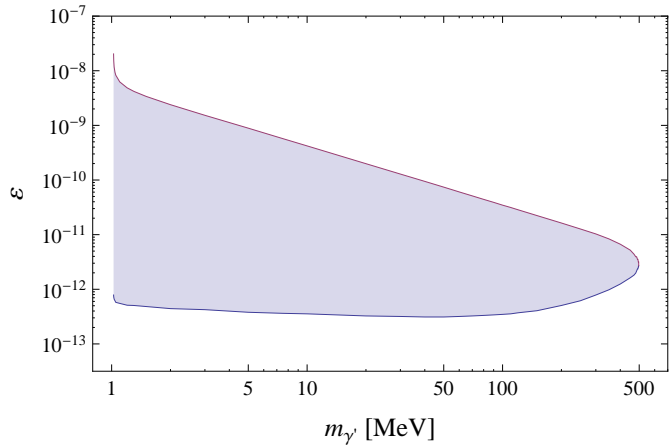}
  \vspace{-.3cm}
  \caption{The shaded region in the $m_{\gamma'} - \epsilon$ plane is excluded by the MeV gamma-ray observations by SMM~\cite{SN-nu1}.}
  \label{fig:constraint_surface}
\end{figure}

The total suppression S of the prompt $X/\gamma$ ray signal from SN1987A is then the product of the above three factors:
\begin{equation}
   S=3\times 10^3 \epsilon^6_{10}
\end{equation}
Recalling that we can tolerate a decrease by at least a factor of $10^{10}$ in the huge benchmark signal that the SMM~\cite{SN-nu1} would have seen had we used $\epsilon_{10}=1$
we infer the following bound:
\begin{equation}
 \epsilon_{10} \leq(1.6 \times 10^{-15 })^{1/6}\sim{3.10^{-3}}.
\end{equation}
This then extends the existing SN cooling limits on $\epsilon$ for almost the full range of the masses by two and a half orders of magnitude, as presented in Fig.~\ref{fig:constraint_surface}. It should be noted that unlike the limits from just the plain cooling arguments our new limits may not improve relative to the higher mass as the $\gamma'$ mass m decreases.

To obtain more precisely the excluded region in Fig.~\ref{fig:constraint_surface}, we first calculate numerically the optical depth as a function of the dark photon mass and mixing parameters
\begin{eqnarray}
\kappa = \frac { (N_{\gamma'} V_{c} \Delta t) R_\ast } { \frac43 \pi (7R_{\ast}^3) } \cdot \sigma_{ann} \,,
\end{eqnarray}
where the factor before the dot is the column number density of the dark photons (and the produced $e^+e^-$ pairs), and when calculating $N_{\gamma'}$ we count only the dark photons decaying within the radius ($R_\ast$, 2$R_\ast$). Then we get the flux of MeV gamma rays
\begin{eqnarray}
\Phi_{\gamma'} = \frac { N_{\gamma'} V_{c} } { 4\pi D^2 } \,.
\end{eqnarray}
Requiring that $\Phi_{\gamma'}$ is less than the SMM bound on the gamma ray background, i.e. $0.1 \, /{\rm cm}^2 /{\rm sec}$, we can exclude some region in the mass-mixing parameter space. In this region, if $\kappa < 1$, the gamma ray flux would be suppressed and it has to be ``rescaled'' by the factor ${\rm min} [1,\, 3\times10^3 \epsilon^2_{10}]$.

Two relevant factors are involved: First the decay length $L_{decay}$ increases for any given ${\gamma'}$ energy larger than $m_{\gamma'}$ in proportion to $E_{\gamma'}/{m^2_{\gamma'}}$, pushing the decays to further out from the star and decreasing the $e^+ e^-$ density and the resulting optical depth for annihilation. The second factor is more subtle and harder to incorporate without a full fledged monte-carlo simulation which we have not attempted here. At any distance $\geq{2R*}\sim{10^7 R_{c}}$ the $\gamma$'s are flowing almost exactly radially outward with the neighboring trajectories being almost exactly parallel. For $E_{\gamma'}/m_{\gamma'} \geq {10-20}$ the trajectories of the electrons and positrons from the $\gamma'$ decays will also be largely parallel and therefore intersect only after large distances, decreasing the annihilation effective cross section by a factor of $\theta=m_{\gamma'}/E_{\gamma'}$ where $\theta$ is the relative lab angle between the trajectories. This may be over compensated by the increase of the annihilation cross section which is proportional
$\sim (E_1E_2\theta^2)^{-2}$ where $E_{1,\,2}$ are the lab energies of the colliding electron and positron. Finally we note that, because of the very strong dependence (as a sixth power) on $\epsilon$, we will not be able to improve the bounds on $\epsilon$ further, even if another more intense and better studied supernova becomes available.

\section{Bounds for the mass range $m_{\gamma'} \leq 2 m_e$}
\begin{figure}
  \centering
  \includegraphics[width=0.5\textwidth]{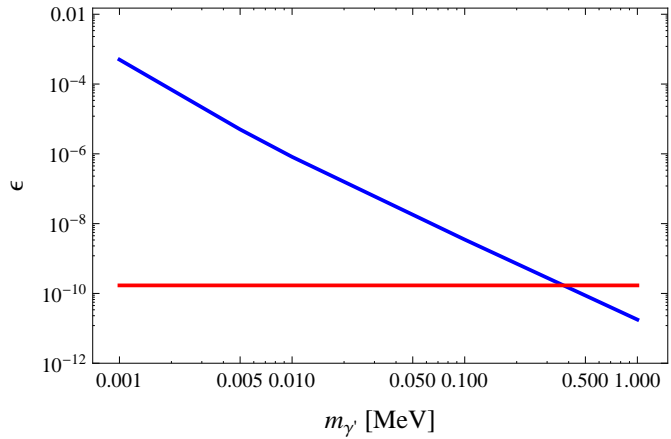}
  \vspace{-.3cm}
  \caption{Constraints on the dark photon parameters for $m_{\gamma'} < 2m_e$, with the red line being the upper bound due to the luminosity, while the blue one the upper bound from the gamma ray considerations at the cosmological scale.}
  \label{fig:constraint_3gamma}
\end{figure}

In this section, we explore the region where the dark photon mass is less than $2m_e$ so that the $e^+e^-$ decay channel is blocked.  The dominant channel in this case is $\gamma' \rightarrow 3 \gamma$ and the decay length of $\gamma'$ far exceeds the progenitor star radius, readily reaching even the Hubble radius. For a 1 MeV dark photon, when $\epsilon_{10} < 2.6$ (assuming $E_{\gamma'} = T$), its lifetime would exceed that of the universe. The first point to note is that the energy loss constraints of course apply to this case. So the question is whether we can improve the bounds from energy loss by using the fact that it decays to photons with a long time scale. The constraint from supernova cooling leads to the horizontal red line given in Fig.~\ref{fig:constraint_3gamma}, independent of the dark photon mass. All the region above this line is excluded.

Since the decay photons will typically have energies of several MeV's and depending on when the Supernova happened (i.e. the red shift at that epoch), there might be some degrading of the energy but nonetheless it is likely to be in the MeV range if we look at the epoch $1+z < 2$. Since there exist severe bounds on the fluence of such photons~\cite{SN-nu1}, we can find constraints on $\epsilon_{10}$.

To derive this constraint, we will consider the cumulative effect of all $\gamma'$'s emitted from all supernovae since $z=1$. The rough estimate for the SN rate/galaxy is one per century. Taking $10^{12}$ galaxies in the universe, we get for the total number of SN after the epoch $z <1$ to be $1.3\times 10^{20}$. The total number of $\gamma'$'s emitted over this time is $\sim 10^{78}\epsilon^2_{10}$. The decay photons energy from the $\gamma'$'s will be visible only if their decay length is less than size of the universe i.e. $L_{\rm decay} < 10^{28}$ cm. The upper bound comes from the fact that the fluence of MeV gamma rays from $\gamma'$ decay is less than the gamma ray background observed by SMM~\cite{SN-nu1}
\begin{eqnarray}
\frac{ 10^{78}\epsilon^2_{10} c}{ 4\pi L_{\rm decay}^3 } < 0.1 {\rm cm}^{-2} {\rm sec^{-1}} \,,
\end{eqnarray}
(where $c$ is the velocity of light) and is presented as the blue line in Fig.~\ref{fig:constraint_3gamma}.

\section{Summary}
\begin{figure}
  \centering
  \includegraphics[width=0.5\textwidth]{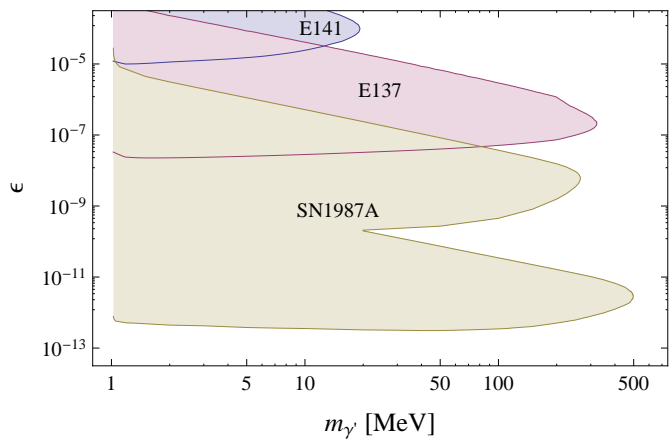}
  \vspace{-.3cm}
  \caption{Parameter space in the dark photon $(m_{\gamma'},\,\epsilon)$ plane excluded by the supernova (relevant) observations and the experiments E137~\cite{dp-ex-E137} and E141~\cite{dp-ex-E141}.}
  \label{fig:exclusion}
\end{figure}

To summarize, we have presented some additional constraints on the photon-dark photon mixing using the possibility that dark photon can decay into $e^+e^-$ after it is emitted from the supernova core if its mass $m > 2 m_e$, using the absence of any prompt MeV range gamma signal shortly after the SN1987A went off. On the other hand, if $m < 2m_e$, its dominant decay mode is to $3\gamma$. This however arises only at the one loop level and is of order $\alpha^4$. In this case, the cumulative effect of all supernovae in the universe would also generate MeV gamma rays. Using the observed luminosity and existing bound on the fluence of such gamma rays, we strengthen the bound on the mixing parameter $\epsilon$ in the smaller mass domain. The constraints for $m_{\gamma'} > 2m_e$ in this work is summarized in Fig.~\ref{fig:exclusion}, which combines the bounds in Fig.~\ref{fig:constraint}, \ref{fig:constraint_mantle} and \ref{fig:constraint_surface}, and exclude large regions of the dark photon mass and mixing parameters, reaching up to $\epsilon_{10} \sim 10^{-2.5}$ for the mass range $2m_e < m_{\gamma'} \lesssim$ 100 MeV. On the other hand, for light dark photons with mass $m_{\gamma'} < 2m_e$, the constraints (presented in Fig.~\ref{fig:constraint_3gamma}) from the radiative decay channel $\gamma' \rightarrow 3\gamma$ exceeds nearly most of the astronomical and cosmological considerations~\cite{dp-review}.

\section*{Acknowledgement}
We would like to thank  Rouven Essig, Maxim Perelstein, Doris C. Rosenbaum, Tod Strohmayer, Nolan R. Walborn, Yue Zhang and Xinmin Zhang for discussion.  Y.Z. thanks the Center for Future High Energy Physics, Institute of High Energy physics, CAS, where this work is finalized and his work is supported in part by the National Natural Science Foundation
of China (NSFC) under Grant No. 11105004. Work of R. N. M. was supported by the NSF grant No. PHY-1315155 and that of DK by NASA ATP and Fermi GI grants. S. N. wishes to thank the Maryland Center for Fundamental Physics for hospitality.

\appendix

\section{Function for the energy emission rate}

Assuming the momenta of the incoming and outgoing particles $N(p_1)N(p_2) \rightarrow N(p_3)N(p_4)\gamma'(p_{\gamma'})$, and redefining the momenta
\begin{eqnarray}
{\bf p}_1 &\equiv& {\bf P} + {\bf p}_i \,, \nonumber \\
{\bf p}_2 &\equiv& {\bf P} - {\bf p}_i \,, \nonumber \\
{\bf p}_3 &\equiv& {\bf P}' + {\bf p}_f \,, \nonumber \\
{\bf p}_4 &\equiv& {\bf P} '- {\bf p}_f \,,
\end{eqnarray}
then the dimensionless parameters are defined as~\cite{dp-ex-SN2}
\begin{eqnarray}
&& u \equiv \frac{{\bf p}_i^2}{m_N T} \,, \nonumber \\
&& v \equiv \frac{{\bf p}_f^2}{m_N T} \,, \nonumber \\
&& x \equiv \frac{E_{\gamma'}}{T} \,, \nonumber \\
&& y \equiv \frac{m_{\pi}^2}{m_N T} \,, \nonumber \\
&& q \equiv \frac{m_{\gamma'}}{T} \,, \nonumber \\
&& z\equiv \cos(\theta_{if}) \,,
\end{eqnarray}
where $\theta_{if}$ is the angle between ${\bf p}_i$ and ${\bf p}_f$.

The function for the $pn \rightarrow pnA'$ production process
\begin{eqnarray}
\mathcal{I}_{pn} = \sum_{i=1}^6 C_i I_i
\end{eqnarray}
where the different pieces
\begin{eqnarray}
I_1 &=& \frac{(u+v-2z\sqrt{uv})^3}{(u+v-2z\sqrt{uv}+y)^2} \,, \nonumber \\
I_2 &=& \frac{(u+v-2z\sqrt{uv})(u+v+2z\sqrt{uv})^2}{(u+v+2z\sqrt{uv}+y)^2} \,, \nonumber \\
I_{3} &=& \frac{ (u+v-2z\sqrt{uv}) (-u^2 -v^2 + (6-4z^2)uv)}{(u+v+y)^2-4z^2uv} \,, \nonumber \\
I_{4} &=& x \frac{(-u^2-v^2+(6-4z^2)uv)}{((u+v+y)^2-4z^2uv)} \frac{(u-v)}{(u+v+y+2z\sqrt{uv} -\frac{T}{m_N}q^2)} \,, \nonumber \\
I_{5} &=& x \frac{(u+v+2z\sqrt{uv})^2}{(u+v+2z\sqrt{uv}+y)^2} \frac{(u-v)}{(u+v+y+2z\sqrt{uv} -\frac{T}{m_N}q^2)} \,, \nonumber \\
I_{6}&=& x^2 \frac{(u+v+2z\sqrt{uv})^3}{(u+v+2z\sqrt{uv}+y)^2} \frac{1}{(u+v+y+2z\sqrt{uv} -\frac{T}{m_N}q^2)^2}
\end{eqnarray}
and the ``coefficients''
\begin{eqnarray}
C_1 &\equiv& 1 \,, \nonumber \\
C_2 &\equiv& 4 \left[ 1 +\frac{ 6x^2 -4 x (u-v) +2q^2}{ (u+v)^2-4z^2uv } \right] \,, \nonumber \\
C_{3} &\equiv& -2 \left[ 1 + \frac{ 2 x(u-v) }{ -u^2-v^2+(6-4z^2)uv } \right] \,, \nonumber \\
C_{4} &\equiv& -4 \,, \nonumber \\
C_{5} &\equiv& 16 \left[ 1 - \frac{2 x}{ u-v } \right] \Rightarrow -16 \,, \nonumber \\
C_{6} &\equiv& 16 \,.
\end{eqnarray}
The delta function $\delta (u-v-x)$ implies that $C_{5} = -16$. The integral is a function of the dimensionless parameters $y$ and $q$, and solely determined by the dark photon mass $m_{\gamma'}$, when the temperature $T$ in the supernova core is fixed.

\end{document}